\documentclass[11pt,a4paper]{article}
\usepackage{jheppub}
\usepackage{graphicx}
\usepackage{epsfig}
\usepackage{amsmath}
\usepackage{amssymb}
\usepackage{amsthm}
\usepackage{ulem}

\def\unit{{1\kern-.65ex {\rm l}}}
\def\1{{1\kern-.65ex {\rm l}}}

\title{A Method for BPS Equations of Vortices}
\author[a,b,1]{A. Nata Atmaja,\note{Corresponding author.}}

\affiliation[a]{Quantum Science Centre, Department of Physics\\
Faculty of Science, University of Malaya\\
50603 Kuala Lumpur, Malaysia.}
\affiliation[b]{Research Center for Physics \\
Indonesian Institute of Sciences (LIPI)\\
Kompleks PUSPITEK Serpong\\
Tangerang 15310, Indonesia.\\}

\emailAdd{ardian\_n\_a@um.edu.my}

\abstract{We develop a new method for obtaining the BPS equations of static vortices motivated by the results of the \textit{On-Shell} method on the standard Maxwell-Higgs model and its Born-Infeld-Higgs model~\cite{Atmaja:2014fha}. Our method relies on the existence of what we shall call an energy function, $Q$, which is a mere function of the (effective) fields. The total energy of BPS vortices, $E_{BPS}$, are simply given by a difference between the boundaries value of $Q$ at $r\to\infty$ and at $r=0$, $E_{BPS}=Q(r\to\infty)-Q(r=0)$. Imposing a condition that these (effective) fields are independent, we may define a BPS Lagrangian, $\mathcal{L}_{BPS}$, derived by taking integral of differential $Q$, $\mathcal{L}_{BPS}=-\int dQ$. Matching the Lagrangian $\mathcal{L}_{BPS}$ with the corresponding effective Lagrangian, we can extract several equations. Solving these equations yields the desired BPS equations and, in some cases, also constraint equations. With our method, the various known BPS equations of vortices are derived in a relatively simple procedure.}

\begin{document}
\maketitle
\flushbottom

\section{Introduction}
In the field theory, vortices are known as point-like solitonic objects in three-dimensional field theory, or extended solitonic objects in higher-dimensional field theory, i.e. vortex strings. In order for vortices to have finite energy, the field theory must be equipped with additional gauge field due to the Derrick's theorem~\cite{Derrick:1964ww}, and so vortices are featured with electromagnetic charges. Vortices find their applications in many branches of physics. As an example, magnetic vortices of the standard Maxwell-Higgs model (sMH), obtained by Nielsen and Olesen~\cite{Nielsen:1973cs}, correspond to the Type-II superconductor identified by Abrikosov~\cite{Abrikosov:1956sx}. Some other applications of vortices are in Bose-Einstein condensates~\cite{PhysRevLett.83.2498}, in quantum Hall effect~\cite{PhysRevLett.62.82}, including cosmic strings in the early formation of the Universe~\cite{Hindmarsh:1994re}, and many more.

The dynamics of vortices is given by the Euler-Lagrange equations of the three-dimensional field theory. This is, however, not an easy task to find solutions that solve the second-order non-linear equations. We could simplify a bit the equations by considering the static vortices and by realizing that vortices are point-like objects hence writing the Euler-Lagrange equations in polar coordinates. In some cases, we may add a constraint or a limit to the parameters of field theory that could make the second-order equations much more simple to solve as the one found by Prasad and Sommerfield in the monopole case~\cite{Prasad:1975kr}. It turns out that these solutions are also solutions to the first-order equations that also solve the second-order equations as shown by Bogomolnyi~\cite{Bogomolny:1975de}. One could obtain these first-order equations directly from the static energy density using Bogomolnyi's trick by squaring the energy density, in which the first-order equations obtained this way are now well-know as Bogomolnyi-Prasad-Sommerfield (BPS) equations. This trick has been used for many solitonic objects including the vortices found in the sMH model~\cite{Nielsen:1973cs}.

The existence of these BPS equations does not only simplify the problem of solving the second-order equations but they also have intimate relation with the supersymmetry extension of the theory~\cite{Weinberg:2012pjx}. Furthermore, in most of the cases, the static energy is bounded from below which is determined fully by topological charge of the theory. The BPS solitons, which are solutions to the BPS equations, saturate this bound of energy thus easy to prove their stability property. However, extracting these BPS equations from the static energy density for general cases is somewhat difficult and tricky. An additional dummy term in the Lagrangian might be needed to execute the Bogomolnyi's trick, with the cost of producing a constraint equation to eliminate this term at the end~\cite{Casana:2015bea,Casana:2015cla}. An attempt was made in~\cite{Bazeia:2007df} by imposing a pressureless condition on the energy-momentum tensor of two-dimensional scalar field theory, with general kinetic term. The kink solutions then can be extracted from this condition. However, in general cases, extracting the BPS equations is also a bit tricky for the higher-dimensional theory. A question was aroused whether there is a more rigorous way to obtain these BPS equations. A proposal was given not along ago and it is know as the \textit{On-Shell} method~\cite{Atmaja:2014fha}. The main idea of this method is by introducing auxiliary fields into the (second-order) Euler-Lagrange equations. These auxiliary fields, which would generate the (first-order) Bogomolnyi equations that are solutions to the Euler-Lagrange equations, are then fixed by splitting the Euler-Lagrange equations into several zeroth-order equations, explicitly independent of space-time coordinates, and then solving these equations. For details procedure see~\cite{Atmaja:2014fha, Atmaja:2015lia}. The advantage of this method is that since one works in the Euler-Lagrange equations, whatever Bogomolnyi equations obtained are always solutions to the Euler-Lagrange equations. However, the procedure is a bit tedious and involves solving more equations than the Euler-Lagrange equations, though they are zeroth-order equations.

In this article, we would like to construct a method that is rigorous and yet as simple as possible to obtaining the BPS equations of vortices. We will apply our method to the standard Maxwell-Higgs model, Born-Infeld-Higgs model, and some variations of those models. The motivation comes from our observation on the results in the static energy formula of BPS vortices derived using the \textit{On-Shell} method~\cite{Atmaja:2014fha}. One of the interesting results of the BPS vortices, obtained using the \textit{On-Shell} method~\cite{Atmaja:2014fha}, was that all static energy of BPS vortices, $E_{BPS}$, are equally written in terms of a function $Q$, which we shall call an energy function. In particular, the static energy $E_{BPS}$ is defined as a difference between the boundaries value of energy function, $Q$, near the boundary, $r\to\infty$ and at the center, $r\to 0$,
\begin{equation}
\label{BPS energy}
 E_{BPS}=Q(r\to\infty)-Q(0)=\int^\infty_0 dQ.
\end{equation}
The energy function of sMH model and its Born-Infeld extension is given by~\cite{Atmaja:2014fha}\footnote{Here we used equal boundaries value for fields $a$ and $f$ on both models; $a(0)=f(\infty)=1$ and $a(\infty)=f(0)=0$.}
\begin{equation}
 \label{Q-function}
 Q=2\pi n~(f^2-1)a,
\end{equation}
where $f$ is the effective complex scalar field, $a$ is the effective gauge field in angular direction, and $n=\pm 1,\pm 2,\pm 3,\ldots$. However, both models require some different fixed form of the scalar potential, $V$, in which for the sMH model is given by~\cite{Nielsen:1973cs,Bogomolny:1975de}
\begin{equation}
 V_{sMH}={1\over 4} (f^2-1)^2,
\end{equation}
while for the Born-Infeld-Higgs (BIH) model is~\cite{Shiraishi:1990zi}
\begin{equation}
 V_{BIH}= b^2\left(1-\sqrt{1-{1\over b^2}(1-f^2)^2}\right),
\end{equation}
with $b$ is related to the overall gauge coupling. This does not come as a surprise since the gauge kinetic terms of both models are different as well. A question arises whether one can find a relation between the gauge kinetic term and the corresponding scalar potential, as such the energy function, $Q$, is given by (\ref{Q-function}), for other models. In more restricted case, all models, with BPS vortices, will have the same energy given that the boundaries value of the fields $f$ and $a$ are also the same. The BPS vortex energy then can be written as in (\ref{BPS energy}).

\section{The construction}
We will start from the energy function $Q$ as given by equation (\ref{Q-function}) and taking the general Abelian-Higgs model to be the following
\begin{equation}
 \mathcal{L}=-K(F_{\mu\nu}F^{\mu\nu})+|D_\mu \phi|^2-V(|\phi|),
\end{equation}
where the signature of the metric is taken to be $(+,-,-)$ and $K,V\ge 0$ are required by unitarity. The covariant derivative is defined as $D_\mu\phi=\partial_\mu \phi+ie A_\mu \phi$ and the field strength is given by $F_{\mu\nu}=\partial_\mu A_\nu -\partial_\nu A_\mu$. We consider the case of static solutions and choose the temporal gauge $A_0=0$. The ansatz for scalar field and gauge field are written in polar coordinates, with spherically symmetric solutions, as follows
\begin{equation}
\label{ansatz}
 \phi=f(r) e^{in\theta}, \qquad \qquad \qquad \qquad A_\theta={1\over e}(a(r)-n),
\end{equation}
where we have chosen a radial gauge $A_r=0$.
Thus
\begin{equation}
\label{kinetic f}
 |D_\mu \phi|^2=-f'^2-\left(af \over r\right)^2,
\end{equation}
and
\begin{equation}
\label{kinetic a}
 B^2\equiv {1\over 2} F_{\mu\nu} F^{\mu\nu}= \left(a' \over r\right)^2.
\end{equation}
Form here now on, we take electromagnetic coupling $e=1$ for simplification. We use a notation that $'\equiv {\partial \over \partial r}$ unless there is an explicit argument on the function, which means taking derivative over the argument.
The effective Lagrangian is now written as
\begin{equation}
\label{eff L-1}
 \mathcal{L}=-K\left(\left(a' \over r\right)^2\right)-f'^2-\left(af \over r\right)^2-V(f).
\end{equation}

For the static case, the energy is related to the Lagrangian as follows
\begin{equation}
 E=-\int d^2x~\mathcal{L}.
\end{equation}
The BPS vortex energy can be written as, using energy function (\ref{Q-function}) and equation (\ref{BPS energy}),
\begin{equation}
E_{BPS}=2\pi \int \left(2fa~df+ (f^2-1)~da\right),
\end{equation}
where we have rescaled the field a in (\ref{Q-function}) to $a/n$, $a\to a/n$, due to the choice of ansatz (\ref{ansatz}). Notes that the fields $a$ and $f$ must be independent of each other.
Realizing both fields $f$ and $a$ depend only on coordinate $r$, we may rewritten the integral on the right hand side of above equation to be the following
\begin{eqnarray}
 \int d^2x~\mathcal{L}_{BPS}&=& 2\pi \int dr~ r\left(-(f^2-1)~{a'\over r}-2fa~{f'\over r}\right)\nonumber\\
 &\equiv& \int d^2x \left(-(f^2-1)~{a'\over r}-2fa~{f'\over r}\right),
\end{eqnarray}
where $\mathcal{L}_{BPS}$ is called BPS Lagrangian and defined as $E_{BPS}=-\int d^2x~ \mathcal{L}_{BPS}$.
So, the BPS Lagragian is simplified to
\begin{eqnarray}
\label{eff L-2}
 \mathcal{L}_{BPS}=-(f^2-1)\left(a'\over r\right)-2\left(af\over r\right)f'.
\end{eqnarray}
As one can see there is a mixing term $f'$ and $af/ r$ on the last term of the BPS Lagrangian (\ref{eff L-2}), while in the second and third terms of effective Lagrangian (\ref{eff L-1}) they are separated. One could think of the effective Lagragian (\ref{eff L-1}) as a sum of BPS Lagrangian (\ref{eff L-2}) and a quadratic terms, which vanishes in the BPS limit. So, it is tempting to have an equation
\begin{equation}
\label{BPS-1}
 2{af\over r}f'=f'^2+\left(af \over r\right)^2.
\end{equation}
 Now, the equation (\ref{BPS-1}) can be rewritten as
\begin{equation}
 \left(f'-{af\over r}\right)^2=0
\end{equation}
and the solution to this equation is given by
\begin{equation}
\label{sol-1}
 rf'=af,
\end{equation}
which is indeed one of the BPS equations for vortices in both models (sMH and BIH). It is easy to show the solution (\ref{sol-1}) solves the equation of motion by rewriting the kinetic terms of scalar field, in Lagrangian (\ref{eff L-1}), as below
\begin{equation}
 -f'^2-\left(af \over r\right)^2=-\left(f'+{af\over r}\right)^2+2{af\over r}f'.
\end{equation}
For the remaining terms, we have
\begin{equation}
\label{BPS-2}
 (f^2-1){a'\over r}=K\left((a'/r)^2\right)+V(f).
\end{equation}
Just like the case of equation (\ref{BPS-1}), solutions to equation (\ref{BPS-2}) can be identified as one of the BPS equations. Similar to the previous one, we expect that
\begin{equation}
\label{BPS-21}
 K\left((a'/r)^2\right)+V(f)=X\left({a'\over r}-Y(f)\right)^p+(f^2-1){a'\over r},
\end{equation}
where $Y$ is a function of $f$, and $p>1$, e.g in the standard case $p=2$. In general $X$ can be an arbitrary function of $f$ and $a'/r$. This is to make sure that the resulting BPS equation, given by $a'/r=Y$, satisfy the Euler-Lagrange equations. The correct function $Y$ should produce the equation (\ref{BPS-21}). A simple way to check whether the chosen function $Y$ is the correct one is by substituting $a'/r=Y(f)$ into the left hand side of equation (\ref{BPS-2}) and check if it is positive definite. However, this simple check is only a necessary condition which may not be correct in general. Another way is by solving equation (\ref{BPS-2}) for $a'/r$ which can be done algebraically. Normally, there will be more than one solutions for $a'/r$ and if this is case, the chosen function $Y$ is the correct one if it equals to at least two solutions of $a'/r$. A more rigorous way to check that a function $Y$ is the correct one can be done by defining a ``semi''-effective Lagrangian
\begin{equation}
 \mathcal{K}\equiv r\left(K+V-(f^2-1){a'\over r}\right).
\end{equation}
and deriving its the Euler-Lagrange equations for $a$ and $f$. If the function $Y$ is the correct one then the substitution of $a'/r=Y$, also using equation (\ref{sol-1}), should solve the corresponding Euler-Lagrange equations. As examples of $Y$ are given by the standard Maxwell-Higgs model, with $K=B^2$, in which $Y=(f^2-1)/2$ and the corresponding BIH model, with
\begin{equation}
 K=b^2\left(\sqrt{1+{1\over b^2}B^2}-1\right),
\end{equation}
in which
\begin{equation}
 Y={f^2-1 \over \sqrt{1-{1\over b^2}(f^2-1)^2}}.
\end{equation}
We will derive them more detail in later sections.

One necessary requirement for the finite energy BPS vortices is that near the boundary, $r\to\infty$, the solutions must reach Higgs vacuum~\cite{Manton:2004tk,Weinberg:2012pjx}. If the real vacuum of the model is defined as the minimum of potential then it implies, in the BPS limit, that the potential must be minimum near the boundary as well. Therefore another way to check if the $Y$ function is the correct one is by analyzing the behavior of potential $V$, in equation (\ref{BPS-2}), near the boundary. Suppose we are given the kinetic term $K$ and it is positive definite in all range of $0\leq r < \infty$. The problem is now to find what solution, or BPS equation $a'/r=Y(f)$, of (\ref{BPS-2}) that give arise to a positive definite $V$ and it has minimum near $r\to\infty$. We can take the solution to be
\begin{equation}
 \left(a'\over r\right)= Y(x),
\end{equation}
where we have defined a new function $x=f^2-1$ such that the equation (\ref{BPS-2}) now becomes
\begin{equation}
\label{pot}
 V(x)=x~Y(x)-K(Y(x))
\end{equation}
and its first and second derivative over $x$ is
\begin{equation}
\label{Dpot}
 V'(x)=Y+\left(x-K'(Y)\right) Y'(x),
\end{equation}
\begin{equation}
\label{DDpot}
 V''(x)=2Y'(x)+\left(x-K'(Y)\right) Y''(x)-K''(Y)\left(Y'\right)^2.
\end{equation}
Notes that, in the BPS limit, $xY$ must be some positive definite function as well, but not necessarily as individual functions $x$ and $Y$, in such away the potential $V$ is also positive definite hence $xY\geq K$ for all value of $x$, or in some allowed range of $x$. This is in accordance with the requirement that the (BPS) vortex energy (\ref{BPS energy}) must be positive definite. Now, assume that near the boundary, $r\to\infty$, $x\to0$ and the minimum potential is zero, $V=0$. To make sure the potential is the minimum there, we require $V'(x)=0$ and $V''(x)>0$. The first condition, $V=0$, yields $K=x~Y$. Now, since the kinetic $K$ must also be zero, so it gives 
\begin{equation}
\label{cond-1}
K=x~Y \longrightarrow Y\geq 0. 
\end{equation}
The second condition, $V'=0$, which can be read from (\ref{Dpot}), and with $x=0$, gives us
\begin{equation}
\label{cond-2}
Y=-\left(x-K'(Y)\right) Y'(x).
\end{equation}
The last condition, $V''(x)>0$, which can also be read from (\ref{DDpot}), yields
\begin{eqnarray}
 \label{cond-3}
 2 Y'(x)+\left(x-K'(Y)\right) Y''(x) &>& K''(Y)Y'(x)^2\nonumber\\
 2 Y'(x)-{Y \over Y'(x) }Y''(x) &>& K''(Y)Y'(x)^2.
\end{eqnarray}
From the last inequality, it is also normal to expect that $Y'(x)\neq 0$ near the boundary.

\section{The standard Maxwell-Higgs model}
The effective action of the standard Maxwell-Higgs model is the following
\begin{equation}
\label{sMH Lagragian}
 \mathcal{L}_{sMH}=-\left(a' \over r\right)^2-f'^2-\left(a~f \over r\right)^2-V(f).
\end{equation}
Now consider an ansatz for energy function 
\begin{equation}
\label{Q-ansatz}
Q=2\pi F(f) A(a), 
\end{equation}
such that the BPS Lagragian is given by
\begin{equation}
\label{BPS Lagragian}
 \mathcal{L}_{BPS}=-F'(f) {A\over r} f'-F~ A'(a) {a'\over r}.
\end{equation}
Matching both these Lagrangians, we may have two equations
\begin{equation}
\label{sMH-BPS 1}
 F'(f) {A\over r} f'=f'^2+\left(a~f \over r\right)^2
\end{equation}
and
\begin{equation}
\label{sMH-BPS 2}
 F~ A'(a) {a'\over r}=\left(a' \over r\right)^2 + V.
\end{equation}
The first equation (\ref{sMH-BPS 1}) can be rewritten as
\begin{equation}
 \left(f'\mp {a f\over r}\right)^2=0.
\end{equation}
Here we have identified $F'(f) A=\pm 2af$ that implies $A=a$ and $F=\pm(f^2+c_F)$, where $c_F$ is a constant. The solutions of the first equation (\ref{sMH-BPS 1}) are given by
\begin{equation}
 \label{BPS equation selfdual}
 f'=\pm {a f\over r},
\end{equation}
which is found in many the literature about BPS vortices. We will call the equation (\ref{BPS equation selfdual}) as a self-dual condition\footnote{This name was introduced by Casana, as informed by Eduardo to us.}. In this article we focus on BPS vortices that bear the self-dual condition (\ref{BPS equation selfdual}) in their BPS equations. We apply the similar steps to the second equation (\ref{sMH-BPS 2}) from which we obtain
\begin{equation}
 \left({a'\over r}\mp\sqrt{V}\right)^2=0.
\end{equation}
The identification is $2\sqrt{V}=(f^2+c_F)$, and so the solutions of the second equation (\ref{sMH-BPS 2}) are given by
\begin{equation}
 a'=\pm{r\over 2}(f^2+c_F).
\end{equation}
If the potential is zero, $V=0$, then one can immediately find the solutions of the second equation (\ref{sMH-BPS 2}),
\begin{equation}
\label{BPS equation sMH V=0}
 a'=\pm r(f^2+c_F),
\end{equation}
which is double that of the non-zero potential's solution. It is also easy to check that $Fa'/r$ is positive definite. Unfortunately, one can check that the BPS equations (\ref{BPS equation selfdual}) and (\ref{BPS equation sMH V=0}) are not solutions of the Euler-Lagrange equations derived from the ``semi''-effective Lagragian
\begin{equation}
 \mathcal{K}=r\left(\left(a' \over r\right)^2-F {a'\over r}\right).
\end{equation}
It is easy to show that the Euler-Lagrange equation for $a$ implies $a'/r=F/2$ which is inconsistent with the solutions (\ref{BPS equation sMH V=0}).

\section{Born-Infeld-Higgs model}
In case of Born-Infeld-Higgs model, the effective Lagrangian can be written by~\cite{Shiraishi:1990zi}
\begin{equation}
\label{BIH Lagragian}
 \mathcal{L}_{BIH}=-b^2\left(\sqrt{1+{1\over b^2}\left(a'\over r\right)^2}-1\right)-f'^2-\left(a f\over r\right)^2-V(f),
\end{equation}
where $b$ is the overall gauge coupling. The matching between BPS Lagrangian (\ref{BPS Lagragian}), obtained using the same ansatz (\ref{Q-ansatz}) for the energy function $Q$, and the Lagragian (\ref{BIH Lagragian}) produces two equations
\begin{equation}
 \label{BIH-BPS 1}
 F'(f) {A\over r} f'=f'^2+\left(a~f \over r\right)^2
\end{equation}
and
\begin{equation}
 \label{BIH-BPS 2}
 F~ A'(a) {a'\over r}=b^2\left(\sqrt{1+{1\over b^2}\left(a'\over r\right)^2}-1\right) + V.
\end{equation}
Similar to the case of standard Maxwell-Higgs model, the first equation (\ref{BIH-BPS 1}) implies $A=a$ and $F=\pm(f^2+c_F)$. The second equation (\ref{BIH-BPS 2}) is hard to be rewritten into quadratic form as in the second equation (\ref{sMH-BPS 2}) of the standard Maxwell-Higgs model. Therefore, we have to solve the equation (\ref{BIH-BPS 2}) for $a'/r$ explicitly. To simplify the problem let us pick $F=f^2-1$, and the solutions are given by
\begin{equation}
 \left(a'\over r\right)_\pm={(f^2-1)(b^2-V)\pm \sqrt{b^4(f^2-1)^2+b^2V(V-2b^2)}\over b^2-(f^2-1)^2}.
\end{equation}
Taking these two solutions to be equal induces solutions for $V$,\footnote{Notes that the solutions of $a'/r$ are not related to solutions of $V$ even though they use the same subscript index $\pm$.}
\begin{equation}
\left(a'\over r\right)_+ = \left(a'\over r\right)_- \longrightarrow
 V_{\pm}=b^2\left(1\pm\sqrt{1-{1\over b^2}(f^2-1)^2}\right).
\end{equation}
The positive solution, $V_+$, is not suggested since there is no global minimum, while the negative solution, $V_-$ is the one obtained in~\cite{Shiraishi:1990zi}. Taking $V=V_-$, the remaining BPS equation is
\begin{equation}
 {a'\over r}={f^2-1 \over \sqrt{1-{1\over b^2}(1-f^2)^2}}.
\end{equation}
One can check that $(f^2-1)a'/r$, in the BPS limit, is not positive definite. However, one should notice that the potential $V_-$ is only positive definite if inequality $b^2\geq (f^2-1)^2$ is satisfied. This inequality also restricts the function $(f^2-1)a'/r$ to be positive definite. One could also check that the resulting BPS equations are solutions to Euler-Lagrange equations for $a$ and $r$ derived from the ``semi''-effective Lagragian
\begin{equation}
 \mathcal{K}=r\left(b^2\left(\sqrt{1+{1\over b^2}\left(a' \over r\right)^2}-1\right)-(f^2-1) {a'\over r}\right).
\end{equation}

Now, suppose we take the potential $V=0$ then the second equation (\ref{BIH-BPS 2}) becomes
\begin{equation}
 (f^2-1) {a'\over r}=b^2\left(\sqrt{1+{1\over b^2}\left(a'\over r\right)^2}-1\right).
\end{equation}
The solutions are given by
\begin{equation}
 {a'\over r}=0,\qquad \qquad \qquad {a'\over r}=2 {b^2 (f^2-1) \over b^2-(f^2-1)^2}.
\end{equation}
The first solution can be neglected. For the second solution, a simple check shows that $(1-f^2)a'/r$ can not be positive definite. So, like the standard Maxwell-Higgs model here the BPS equations of BIH vortex with zero-potential do not satisfy the Euler-Lagrange equations. 

\subsection{Extension with a non-linear coupling and zero potential}
An extension of the Born-Infeld-Higgs model was discussed in~\cite{Shiraishi:1990zi} by introducing a non-linear coupling between the gauge and Higgs fields and setting the potential to zero. This model has a close connection with the supersymmetric extension of Born-Infeld-Higgs (\ref{BIH Lagragian}), studied in~\cite{Deser:1980ck}, and its effective action is given by
\begin{equation}
 \label{BIH-Ext Lagragian}
 \mathcal{L}_{Ext}=-b^2\left(G(f)\sqrt{1+{1\over b^2}\left(a'\over r\right)^2}-1\right)-f'^2-\left(a f\over r\right)^2,
\end{equation}
where $G>0$ is coupling function of $f$.
In this case the second equation (\ref{BIH-BPS 2}) turns to
\begin{equation}
 F {a'\over r}=b^2\left(G\sqrt{1+{1\over b^2}\left(a'\over r\right)^2}-1\right).
\end{equation}
The solutions for $a'/r$ of this equation are
\begin{equation}
 \left(a' \over r\right)_{\pm}={b^2F \pm b^2 G \sqrt{F^2-b^2 G^2+ b^2} \over b^2 G^2-F^2}.
\end{equation}
Equating these two solutions yields solutions for $G$,
\begin{equation}
 G_1=0,\qquad \qquad \qquad G_2=-\sqrt{1+{F^2 \over b^2}},\qquad \qquad \qquad G_3=\sqrt{1+{F^2 \over b^2}}
\end{equation}
Thus the corect solution for $G$ is $G_3=\sqrt{1+{1 \over b^2}(f^2+c_F)^2}$, which is equal to the one obtained in~\cite{Shiraishi:1990zi}.

\subsection{Generalized Born-Infeld-Higgs model}
The generalized Born-Infeld-Higgs model was proposed in~\cite{Casana:2015bea}. The corresponding effective action is
\begin{equation}
\label{Gen BIH Lagrangian}
 \mathcal{L}_{GBIH}=-b^2\left(\sqrt{1+{G(f)\over b^2}\left(a'\over r\right)^2}-1\right)-w(f)\left(f'^2+\left(a f\over r\right)^2\right)-V(f).
\end{equation}
Matching the above effective Lagragian with the BPS Lagrangian (\ref{BPS Lagragian}), using the ansatz (\ref{ansatz}), yields the following equations
\begin{equation}
\label{GBIH-BPS 1}
 F'(f) {A\over r} f'=w\left(f'^2+\left(af \over r\right)^2\right)
\end{equation}
and
\begin{equation}
\label{GBIH-BPS 2}
 F {a'\over r}=b^2\left(\sqrt{1+{G\over b^2}\left(a'\over r\right)^2}-1\right)+V.
\end{equation}
The first equation (\ref{GBIH-BPS 1}) has the self-dual condition (\ref{BPS equation selfdual}) as its solution with identifications $A=a$ and $F'(f)=\pm 2f w$. Now, the solutions to the second equation (\ref{GBIH-BPS 2}) are
\begin{equation}
 \left(a' \over r\right)_\pm = {F(V-b^2)\pm\sqrt{b^2V^2G+b^4(F^2-2 G V)} \over F^2-b^2 G^2}.
\end{equation}
Equating these solutions implies solutions for $F$,
\begin{equation}
 F=\pm b^2\sqrt{{GV \over b^2}\left(2b^2-V\right)}.
\end{equation}
Using notation in~\cite{Casana:2015bea}, we change $V\to b^2(1-V)$ and so the constraint equation becomes
\begin{equation}
 F'(f)=\pm{\partial \sqrt{b^2 G(1-V^2)}\over \partial f}=\pm 2 w f.
\end{equation}
This constraint equation, however, is different from the one in~\cite{Casana:2015bea} by a negative sign. This is due to different definition of covariant derivative used in~\cite{Casana:2015bea}, $D_\mu=\partial_\mu-ie A_\mu$. And notice that the right hand side of constraint equation will have a multiplication factor of $e$ if we write the electromagnetic coupling explicitly.

\section{Generalized Maxwell-Higgs model}
For the generalized Maxwell-Higgs model, the effective Lagrangian is~\cite{Bazeia:2012uc}
\begin{equation}
\label{Gen Lagrangian}
 \mathcal{L}_G=-{G(f)\over 2}\left(a' \over r\right)^2-w(f)\left(f'^2+\left(af \over r\right)^2\right)-V(f).
\end{equation}
Again matching with the BPS Lagrangian (\ref{BPS Lagragian}) yields the similar first equation (\ref{GBIH-BPS 1}). So, it produces the self-dual condition (\ref{BPS equation selfdual}) and a constraint $F'(f)=\pm 2 w f$, with $A=a$. The second equation is given by
\begin{equation}
\label{Gen-BPS 2}
F A'(a){a' \over r}={G\over 2}\left(a' \over r\right)^2+V,
\end{equation}
and it can be rewritten as
\begin{equation}
 {G\over 2}\left({a' \over r}\mp \sqrt{2V \over G}\right)^2=0,
\end{equation}
with an identification
\begin{equation}
 F=\pm\sqrt{2VG}.
\end{equation}
Thus the BPS equation is
\begin{eqnarray}
\label{BPS equation C_0=0}
 {a' \over r}&=&\pm \sqrt{2V \over G},
\end{eqnarray}
with a constraint equation
\begin{equation}
 {\partial F \over \partial f}=\pm{\partial(\sqrt{2VG}) \over \partial f}=\pm 2wf.
\end{equation}
So, we reproduce the BPS equations, (\ref{BPS equation selfdual}) and (\ref{BPS equation C_0=0}), and a constraint ${\partial(\sqrt{2VG}) \over \partial f}= 2wf$ of the BPS vortices that of $C_0=0$ type in~\cite{Atmaja:2015lia}, also~\cite{Bazeia:2012uc}.


\section{{\it K}-generalized Abelian-Higgs model}
The general form of {\it k}-generalized Abelian-Higgs model has been proposed recently in~\cite{Casana:2015cla}, in which the Lagrangian is given by
\begin{equation}
 \mathcal{L}_{k}=h(|\phi|) K\left(- {F_{\mu\nu}F^{\mu\nu} \over 4~\mathcal{U}(|\phi|)}\right)+w(|\phi|)|D_\mu\phi|^2-V(|\phi|).
\end{equation}
Using the ansatz (\ref{ansatz}) the effective Lagrangian is
\begin{equation}
\label{k-Gen Lagrangian 1}
 \mathcal{L}_{k}=h(f) K\left(- {a'^2 \over 2r^2\mathcal{U}(f)}\right)-w(f)\left(f'^2+\left(a f\over r\right)^2\right)-V(f).
\end{equation}
As before, we take the ansatz (\ref{Q-ansatz}) to arrive to the BPS Lagrangian (\ref{BPS Lagragian}). Comparing the effective Lagrangian (\ref{k-Gen Lagrangian 1}) with the BPS Lagrangian (\ref{BPS Lagragian}) implies two equations :
\begin{equation}
\label{k-Gen-BPS 1}
 F'(f) {A\over r} f'=w(f)\left(f'^2+\left(a f\over r\right)^2\right)
\end{equation}
and 
\begin{equation}
\label{k-Gen-BPS 2}
 F~A'(a) {a' \over r}= V-h~ K\left(- {a'^2 \over 2r^2\mathcal{U}}\right).
\end{equation}
The first equation (\ref{k-Gen-BPS 1}) will give us, as usual, the self-dual condition (\ref{BPS equation selfdual}) and a constraint $F'(f)=\pm 2wf$, with $A=a$. The second equation (\ref{k-Gen-BPS 2}) is rather difficult to solve unless we know the explicit form of $K$. In principal, one should get solutions for $a'/r$ as functions of $f$.

\subsection{New self-dual solutions}
\label{First derivation}
One way to solve this without knowing the explicit form of the kinetic term $K$ is by introducing new terms into the Lagragian as suggested in~\cite{Casana:2015cla} as follows
\begin{equation}
\label{k-Gen Lagrangian 2}
 \mathcal{L}_k=-{1\over 2}\left(K_\mathcal{F} B^2+{W^2\over K_\mathcal{F}}\right)+w|D_\mu\phi|^2+h~K+{1\over 2}\left(K_\mathcal{F} B^2+{W^2\over K_\mathcal{F}}\right)-V,
\end{equation}
where $W$ is a function of scalar field and $K_\mathcal{F}\equiv {\partial K \over \partial \mathcal{F}}$, with $\mathcal{F}=-B^2/2\mathcal{U}$. There is no change to the first equation (\ref{k-Gen-BPS 1}). However, now the first two terms in the Lagragian (\ref{k-Gen Lagrangian 2}) contribute to the right hand side of second equation (\ref{k-Gen-BPS 2}) in which now becomes
\begin{equation}
\label{k-Gen-BPS 3}
F \left(a'\over r\right)= {1\over 2}\left(K_\mathcal{F} \left(a'\over r\right)^2+{W^2\over K_\mathcal{F}}\right).
\end{equation}
We can rewrite the new second equation (\ref{k-Gen-BPS 3}) to
\begin{equation}
 {K_\mathcal{F} \over 2}\left({a'\over r}\mp {W\over K_\mathcal{F}}\right)^2=0,
\end{equation}
where we have identified $F=\pm W$. Thus we obtain a BPS equation
\begin{equation}
\label{BPS equation k-Gen 1}
 {a'\over r}=\pm {W\over K_\mathcal{F}}
\end{equation}
and a constraint equation
\begin{equation}
\label{k-Gen constraint 1}
{\partial W \over \partial f}= 2w f. 
\end{equation}
The remaining terms of Lagragian (\ref{k-Gen Lagrangian 2}) gives another constraint equation
\begin{equation}
 h~K+{1\over 2}\left(K_\mathcal{F} B^2+{W^2\over K_\mathcal{F}}\right)=V.
\end{equation}
Using the BPS equation (\ref{BPS equation k-Gen 1}), it can be simplified to
\begin{equation}
\label{k-Gen constraint 2}
 h~K=V-{W^2 \over K_\mathcal{F}}=V-K_\mathcal{F} B^2.
\end{equation}

A simple case was considered in~\cite{Casana:2015cla} by taking $w=1$ in which, from the constraint (\ref{k-Gen constraint 1}), implies $W=f^2+c_W$, where $c_W$ is a constant. Further, they considered $V=0$ and $h=\mathcal{U}$, in which the BPS equation (\ref{BPS equation k-Gen 1}) now becomes
\begin{equation}
\label{BPS equation k-Gen 2}
{a'\over r} = \pm {f^2+c_W \over K_\mathcal{F}}
\end{equation}
and the constraint equation (\ref{k-Gen constraint 2}) is now simplified to
\begin{equation}
\label{k-Gen constraint 3}
 \mathcal{U}~K= -K_\mathcal{F}B^2,
\end{equation}
or
\begin{equation}
\label{k-Gen constraint 4}
 K_\mathcal{F}={1\over 2 \mathcal{F}}K.
\end{equation}
It is somehow peculiar and seems to restrict the function $K$ that can satisfy the above constraint equation. However, in order for arbitrary $K<0$ to satisfy the above constraint equation then $\mathcal{F}<0$ must be a c-number, with $K_\mathcal{F}>0$, since the above constraint equation only depends on $\mathcal{F}$.

\subsection{Alternative derivation}
\label{Second derivation}
One could actually simplify the previous procedures by assuming the kinetic terms of Lagragian (\ref{k-Gen Lagrangian 1}) can be written as the right hand side of equation (\ref{k-Gen constraint 3}), with $h=\mathcal{U}$, as follows
\begin{equation}
\label{k-Gen Lagrangian 3}
 \mathcal{L}_k=- K_\mathcal{F}\left(B,f\right)~B^2-w(f)\left(f'^2+\left(a f\over r\right)^2\right)-V(f).
\end{equation}
Notes here in general $K_\mathcal{F}$ has nothing to do with $K$. Using the same ansatz for (\ref{Q-ansatz}) and matching the effective Lagrangian (\ref{k-Gen Lagrangian 3}) with the BPS Lagragian (\ref{BPS Lagragian}) yields, again, the first equation (\ref{k-Gen-BPS 1}) which implies the self-dual condition (\ref{BPS equation selfdual}) and a constraint $F'(f)=\pm 2wf$, with $A=a$. The second equation is now given by
\begin{equation}
\label{k-Gen-BPS 4}
 F {a'\over r}=K_\mathcal{F}(B^2,f) \left(a'\over r\right)^2+V
\end{equation}
and it can be rewritten as
\begin{equation}
 K_\mathcal{F} \left({a'\over r}\mp\sqrt{V \over K_\mathcal{F}}\right)^2=0,
\end{equation}
where we have identified $F=\pm 2\sqrt{V~K_\mathcal{F}}$. So, we obtain a BPS equation
\begin{equation}
\label{BPS equation k-Gen 3}
 {a'\over r}=\pm\sqrt{V \over K_\mathcal{F}}
\end{equation}
and a constraint equation
\begin{equation}
\label{k-Gen constraint 5}
 {\partial\left(\sqrt{V~K_\mathcal{F}}\right) \over \partial f}= w f.
\end{equation}
The above constraint equation (\ref{k-Gen constraint 5}) is very complicated and doe not seem to make sense since $K_\mathcal{F}$ in general may be a function of $B^2=(a'/r)^2$. We may substitute $B^2$ using the BPS equation (\ref{BPS equation k-Gen 3}), but, before doing that, we need to find the solutions of BPS equation (\ref{BPS equation k-Gen 3}) for $a'/r$ which are functions of $f$ only. Then we need to pick ones of those solutions that also solve the Euler-Lagrange equations of the following ``semi''-effective Lagragian
\begin{equation}
 \mathcal{K}=r\left(K_\mathcal{F}\left(a'\over r\right)^2+V-F {a'\over r}\right).
\end{equation}

Now consider the case where the potential $V=0$, and so the second equation (\ref{k-Gen-BPS 4}) is simplified to
\begin{equation}
  F {a'\over r}=K_\mathcal{F}(B^2,f) \left(a'\over r\right)^2.
\end{equation}
One can immediately obtain the BPS equation
\begin{equation}
\label{BPS equation k-Gen 4}
 {a'\over r}={F\over K_\mathcal{F}},
\end{equation}
which is equal to the BPS equation (\ref{BPS equation k-Gen 2}) if $w=1$, $F'(f)=\pm 2f$. Here, we reproduce the results in subsection \ref{First derivation} by identifying $K_\mathcal{F}\equiv \partial K/ \partial\mathcal{F}$. 
However, one need to solve the BPS equation (\ref{BPS equation k-Gen 4}) for $a'/r$ and check if it satisfies the Euler-Lagrange equations of the following ``semi''-effective Lagrangian
\begin{equation}
 \mathcal{K}=r\left(K_\mathcal{F} \left(a'\over r\right)^2-F {a'\over r}\right).
\end{equation}
The Euler-Lagrange equation for $a$ is
\begin{equation}
 F=\left(2 K_\mathcal{F}+a' {\partial K_\mathcal{F} \over \partial (a')}\right){a' \over r}.
\end{equation}
Taking that $K_\mathcal{F}\equiv K_\mathcal{F}(\mathcal{F}(a'/r,f))$, it can be rewritten as
\begin{equation}
 F=2\left( K'(\mathcal{F})+\mathcal{F} K''(\mathcal{F})\right){a' \over r}.
\end{equation}
Notice that the above equations are without taking the BPS limit. Comparing the last equation, for $a$, with the BPS equation (\ref{BPS equation k-Gen 4}) implies a constraint
\begin{equation}
 2 K''(\mathcal{F})\mathcal{F}+K'(\mathcal{F})=0,
\end{equation}
which is another way of writing the constraint equation (\ref{k-Gen constraint 4}).

\section{Summary}
The procedure of our method, that we have constructed in this article, can be summarized into the following steps:
\begin{enumerate}
 \item Write the effective Lagrangian, $\mathcal{L}_{eff}$, in terms of effective fields of the model, $\phi_i$ for $i=1,\ldots,N$, by imposing some ansatz.
 
 \item Take an ansatz for energy function as function of the effective fields, $Q\equiv Q(\vec{\phi})$, where $\vec{\phi}\equiv (\phi_1,\ldots,\phi_N)$. In most of the cases, the energy function is written as separable functions of the effective fields, $Q\equiv \prod^N_{i=1} Q_i(\phi_i)$.
 
 \item Derive the BPS Lagrangian, $\mathcal{L}_{BPS}$, from the energy function. Notice that the effective fields are independent of each others such that 
 \[\mathcal{L}_{BPS}=-\int \sum^N_{i=1} Q'(\phi_i) d\phi_i.\]
 
 \item Extract several equations from matching the effective Lagragian with the BPS Lagrangian, $\mathcal{L}_{eff}=\mathcal{L}_{BPS}$. The number of these equations must be equal with the number of effective fields and, in most of the cases, each of these equations does not contain more than one type of first-derivative effective fields,
 \[
  \Phi_i(\phi_i',\vec{\phi},r)=0,
 \]
 with $i=1,\ldots,N$.
 
 \item Find the solutions of these equations for each first-derivative effective fields, $\phi_i'$. Along with these solutions, sometimes one has to make some identifications which later become constraint equations.
 
 \item Pick the solutions for $\phi_i'$ that satisfy the Euler-Lagrange equations of the effective Lagrangian. The corresponding solutions can be identified as the BPS equations. A simple way to do this is by equating two of solutions to $\phi_i'$. Another way is by defining ``semi''-effective Lagrangians, $\mathcal{K}_i$, from $\Phi_i$, and then derive the corresponding Euler-Lagrange equations. The desired solution for $\phi_i'$ must solve these Euler-Lagrange equations.
\end{enumerate}

Here, we have applied our method to various models and obtained the BPS equations of vortices with a relatively simple procedure, compared to the \textit{On-Shell} method~\cite{Atmaja:2014fha, Atmaja:2015lia}. All of the models that are discussed in this article have two general features. First, the self-dual condition (\ref{BPS equation selfdual}) is always be one of the BPS equations. Second, the energy function has a general form of
\begin{equation}
 Q=2\pi a F(f),
\end{equation}
where $F'(f)=2 w f$, and $w$ is a function of $f$ interpreted as an overall coupling of the kinetic terms of the scalar field.

However, our method has a drawback since we must rely on the existence of the energy function in the first place. There are probably some vortices with energy that can not be written merely in terms of the energy function, $Q$. As an example is the $C_0\neq 0$ type vortices of the generalized Maxwell-Higgs model obtained in ~\cite{Atmaja:2015lia}. Nevertheless, we expect that our method here can also be used to find other BPS vortices, such as in the Chern-Simon-Higgs models, and furthermore to find the BPS equations of solitons in the higher-dimensional field theory.

Possible variations can be made to our method. As an example, one can extract the following equations, from matching the sMH Lagrangian (\ref{sMH Lagragian}) and the BPS Lagragian (\ref{BPS Lagragian}),
\begin{equation}
 F'(f) {A\over r} f'=f'^2+V
\end{equation}
and 
\begin{equation}
 F A'(a) {a'\over r}=\left(a' \over r\right)^2+\left(a f  \over r\right)^2.
\end{equation}
From these equations, one may obtain the BPS equations
\begin{equation}
 f'=\pm \sqrt{V}
\end{equation}
and
\begin{equation}
 a'=\pm a f
\end{equation}
along with some identifications: $A=a$ and $FF'(f)=4f\sqrt{V}$. These identifications imply that function $a/r$ is function of $f$, which does not violate our assumption that $a$ and $f$ are independent. It is interesting to notes that the self-duality condition (\ref{BPS equation selfdual}) is not part of the BPS equations above. Nevertheless, one has to check further if these BPS equations and the identifications are consistent and the later whether the solutions do exist or not, which are beyond the discussions in this article.

Separation variables in the energy function, $Q$, might be related with the separation variables in the auxiliary fields introduced in the \textit{On-Shell} method. One may try to relax this condition by allowing the energy function not be separable, e.g. $Q=F(f)+A(a)$. Another thing that one can try if by allowing the energy function to depend explicitly on coordinates, which in this article is only the radial coordinate, such as $Q\equiv Q(a,f,r)$. These variations might be needed to find the BPS equations of vortices of the models that are not discussed in this article.

\acknowledgments
We would like to acknowledge University of Malaya for the support through the University of Malaya Research Grant (UMRG) Programme RP006C-13AFR and RP012D-13AFR.

\bibliography{testBiB}
\bibliographystyle{hieeetr}
\end{document}